\documentclass[prl,aps]{revtex4}

\usepackage{bm}
\usepackage{graphicx}
\usepackage{amsfonts}
\usepackage{amsmath}
\usepackage{amssymb}
\newcommand {\be}{\begin{equation}}
\newcommand {\ee}{\end{equation}}
\newcommand {\bea}{\begin{eqnarray}}
\newcommand {\eea}{\end{eqnarray}}

\begin{document}

\title{Electromagnetic response of broken-symmetry nano-scale clusters}

\author{Ilya Grigorenko}
\affiliation{Department of Physics and Astronomy, University of Southern
California, Los Angeles, CA 90089-0484}
\author{Stephan Haas}
\affiliation{Department of Physics and Astronomy, University of Southern
California, Los Angeles, CA 90089-0484}
\author{A.F.J. Levi}
\affiliation{Department of Physics and Astronomy, University of
Southern California, Los Angeles, CA 90089-0484}
\affiliation{Department of Electrical Engineering, University of
Southern California, Los Angeles, CA 90089-2533}

\date{\today}

\begin{abstract}
A microscopic, non-local response theory is developed to model the
interaction of electromagnetic radiation with inhomogeneous
nano-scale clusters. The breakdown of classical continuum-field
Mie theory is demonstrated at a critical coarse-graining
threshold, below which macroscopic plasmon resonances are replaced
by molecular excitations with suppressed spectral intensity.
\end{abstract}


\maketitle

The prevalent classical model describing the interaction of
visible and infrared electromagnetic radiation with nano-scale
metallic clusters is based on Mie theory \cite{Mie}. This local
continuum field model which employs empirical values of a bulk
material's linear optical response has been used to describe
plasmon resonances in nanoparticles
\cite{Ashcroft,DasSarma,Fredkin}. However, such a semi-empirical
continuum description necessarily breaks down beyond a certain
level of coarseness introduced by atomic length scales. Thus, it
cannot be used to describe the interface between quantum and
macroscopic regimes. Moreover, extensions of Mie theory to
inhomogeneous cluster shapes are commonly restricted to low order
harmonic expansions (e.g. elliptical distortions) and so do not
exhaust the full realm of possible structures. In addition,
near-field applications, such as surface enhanced Raman scattering
\cite{sers}, are most naturally described using a real-space
theory that includes the {\it non-local} electronic response of
inhomogeneous structures, again beyond the scope of Mie theory.

Ignoring the limitations of Mie theory can lead to conclusions of
limited validity and usefulness. For example, it was recently
argued \cite{Fredkin} that  resonances in the long wavelength
limit are exclusively controlled by the object's shape, but
otherwise {\it scale-invariant}, so that scaling of nanoparticles
would not alter their characteristic plasmon frequencies. In
particular, it was stated that an infinitely long elliptic rod,
whose surface in Cartesian coordinates is defined by $(x/a)^2 +
(y/b)^2 = R^2$, has two geometric {\it scalable} plasmon
resonances that depend only on the ratio between semi-axis lengths
$a$ and $b$. In this work we develop a microscopic model that
demonstrates the breakdown of this concept at atomic scales,
whereas for large cluster sizes the classical predictions for the
plasmon resonances are reproduced. We then show how this approach
is useful in describing the coexistence of excitations of quantum
and classical character
 in inhomogeneous structures on the nanoscale.

To capture the main single-particle and collective aspects of
light-matter interaction  in an inhomogeneous nanoscale system
 we adopt the well established linear response
approximation \cite{whaley1}. Starting from the Schr\"odinger
equation for noninteracting electrons with mass $m_{\text e}$ and
charge $e$ moving in
 potential $V({\bf r})$ we have
\begin{eqnarray}
\label{eq1} \label{hamiltonian} H \Psi_i({\bf r})=
\left(-\frac{\hbar^2}{2 m_{\text e}}\nabla^2+V({\bf
r})\right)\Psi_i({\bf r}) = E_i \Psi_i({\bf r}).
\end{eqnarray}
Eq. \ref{eq1} is solved simultaneously with the Poisson equation
that determines the local potential due to the spatial
distribution of the positive background charges. Using the jellium
approximation, the resulting potential is implicitly given by
$\nabla^2 V({\bf r})=4 e \pi \rho({\bf r})$, where the density of
the positive background charge $\rho({\bf r})$ satisfies the
condition of neutrality inside the nanostructure so that $ \int
\rho({\bf r})d {\bf r}=N_{\text {el}}$, where $N_{\text {el}}$ is
the number of electrons.

Within  linear response theory  \cite{Lindhard,RPA}, the induced
electron charge density due to an external field of frequency
$\omega$ is given by $\rho _{\text {ind}} ({\bf r},\omega) = \int
{\chi ({\bf r},{\bf r'},\omega ) \phi _{\text {tot}} ({\bf
r'},\omega)
 }d {\bf r'}$,
where the total self-consistent potential is determined by $ \phi
_{{\text {tot}}} ({\bf r'},\omega) = \phi _{{\text {ext}}} ({\bf
r'},\omega) + \phi _{{\text {ind}}} ({\bf r'},\omega) $. Because
the significance of non-local effects for inhomogeneous systems,
including spatial dispersion of the dielectric function,
 is widely recognized \cite{wake-nonlocal,surface}, for an
inhomogeneous cluster of arbitrary shape we use the non-local
density-density response function $\chi ({\bf r},{\bf r'},\omega
)$ that may be determined from a real-space representation
\cite{Ashcroft}. Within the random phase approximation \cite{RPA}
one finds $\chi ({\bf r},{\bf r'},\omega ) = \sum\limits_{i,j}
{\frac{{f(E_i ) - f(E_j )}}{{E_i - E_j  - \hbar \omega  - i\gamma
}}\Psi _i^* ({\bf r})} \Psi _i ({\bf r'})\Psi _j^* ({\bf r'})\Psi
_j^{} ({\bf r})$, where $f(E_i)$ is the Fermi filling factor and
the small constant $\gamma$ describes level broadening. Electron
eigenenergies $E_i$ and eigenfunctions $\Psi_i$ are obtained
numerically using a $6$th order real-space discretization of the
Schr\"odinger equation in three dimensions with a star-like
stencil on a simple cubic lattice. Unless explicitly stated
otherwise,  zero Derichlet boundary conditions are used to
calculate the electron wave functions in this paper. The large
scale sparse eigenproblem is solved using ARPACK \cite{arpack}.
The induced potential $\phi _{\text{ind}}$ is determined from the
self-consistent integral equation \bea \label{integral_equation}
\phi _{{\text {ind}}} ({\bf r},\omega) &=& \sum\limits_{i,j}
{\frac{{f(E_i ) - f(E_j )}}{{E_i  - E_j  - \hbar \omega  - i\gamma
}}} \int {\frac{{\Psi _i^* ({\bf r'})\Psi _j^{} ({\bf
r'})}}{{\left| {{\bf r} - {\bf r'}} \right|}}d{ \bf r'}} \nonumber
\\ &\times& \int {\Psi _j^* ({\bf r''}) \phi _{\text {tot}} ({\bf
r'},\omega)\Psi _i^{} ({\bf r''})d {\bf r''}}, \eea and the
induced field is found via ${\bf E}_{{\text {ind}}}({\bf
r},\omega)=-{\bf {\nabla}} \phi _{{\text {ind}}} ({\bf
r},\omega)$. For sufficiently small structures, when the typical
wavelength of the external radiation is larger than the structure,
retardation effects can be ignored. Note, that the induced field
is not necessarily collinear with the applied external field
${{\bf E}}_{\text {ext}}({\bf r})$.

The integral equation Eq.(\ref{integral_equation}) is discretized
on a real-space mesh, using a simple cubic lattice with lattice
constant $L$. The natural energy scale $E_0$ is defined by
$E_0=\hbar^2/(2 m_{\text e} L^2)$. For example, in the case of
GaAs $1$D nanorods \cite{nanorod} with conduction band carrier
density $10^{18}$ cm$^{-3}$ the discretization lattice length
scale is $L = 1.28$ nm, the density $\rho=1/480$ in units of
$L^{-3}$ and the energy scale  $E_0= 330$ meV.

%The system of linear equations is solved using a conjugated
%gradient technique rather than by calculating the numerically
%expensive inverse of the dielectric matrix $\epsilon({\bf r},{\bf
%r^\prime},\omega)$. In each gradient step the corresponding
%Poisson problem is solved using  an efficient multigrid method
%\cite{multigrid}.
\begin{figure}[h]
\vspace{0.2cm}
\includegraphics[width=5.cm,angle=-90, bb=33 43 570 713]{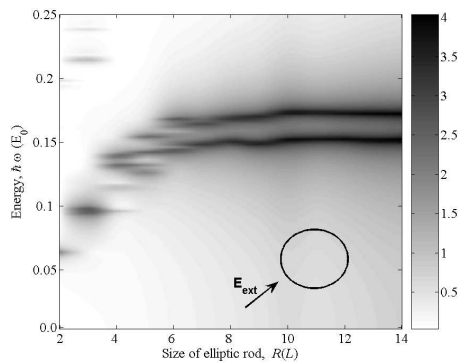}
%\includegraphics[width=7.5cm,angle=270]{tmp2.ps}
%\vspace{-1.5cm}
\vspace{-0.3cm} \caption{\label{fig1}\footnotesize {Logarithm of
the energy of the induced electric field
$(\log_{10}(W_{\text{ind}}))$ (see text) in an elliptic metallic
rod due to applied external field ${\bf E}_{\text{ext}}$ along the
$(1,1,0)$ direction as a function of face surface size $R$ and
photon energy $\hbar \omega$ of the external field in units of
$E_0$.  Aspect ratio (semi-minor to semi-major axis) $a:b=1:1.3$.
Cross section of the elliptical rod and the vector of the applied
external field ${\bf E}_{\text{ext}}$ are depicted in the lower
right corner. Temperature $T=0$ K, carrier density $\rho=(1/480)$
$L^{-3}$, and $\gamma=10^{-3}$ $E_0$. All real space calculations
are performed on a mesh of simple cubic geometry with lattice
constant $L$. }}
\end{figure}
We illustrate the application of this approach to an infinitely
long elliptic rod illuminated by an external field.  The rod is
aligned in the $z$ direction and the electric field polarization
is along the $(1,1,0)$  direction. For the wave functions we
assume periodic boundary conditions along the $z$ direction.
% and in the
%other directions with choose sufficiently large computation region
%with zero boundary conditions.
To quantify the response of arbitrary geometries to the applied
field, we calculate the energy of the induced field, defined by $
W_{\text {ind}}(\omega) = (1/2) \int {\left| {{\bf E}_{\text
{ind}}({\bf r},\omega) } \right|^2 d {\bf r}}$. In Fig. \ref{fig1}
logarithm of the energy of the induced field
$(\log_{10}(W_{\text{ind}}))$ is displayed as a function of the
applied external field photon energy $\hbar \omega$ and the
characteristic system size $R$ for the aspect ratio $a:b=1:1.3$.
For sufficiently large rod sizes, one clearly observes two plasmon
resonances, $\omega_+$ and $\omega_-$, consistent with earlier
predictions based on Mie theory \cite{Fredkin}. Our method
confirms that these resonances occur at $\omega_+ = \omega_{\text
p} (b/(a+b))^{1/2}$ and $\omega_- = \omega_{\text p}
(a/(a+b))^{1/2}$, where $\omega_{\text p}$ is the bulk plasmon
frequency $\omega_{\text p}=\sqrt{4\pi e^2 \rho/ m_{\text {e}}}$.
The grey scale in Fig. \ref{fig1} is on a logarithmic scale,
demonstrating that the spectral intensity for large rod sizes is
orders of magnitude greater compared to smaller rod sizes. Most
importantly, it is evident from this figure that the classical
picture of two well-defined resonances breaks down below a
characteristic system size. For sufficiently small rod sizes, the
two macroscopic resonances split into multi-level molecular
excitations, with the overall spectral weight shifting towards
lower energies. For the chosen parameters this transition occurs
at $R_{\text {c}} \approx 6.5$ $L$. Below $R_{\text {c}}$, the
spectral intensity of the energy levels is
 reduced because fewer electrons
participate in the individual resonances.
\begin{figure}[h]
\vspace{0.cm}
\includegraphics[width=5.cm,angle=270, bb=50 71 559 735]{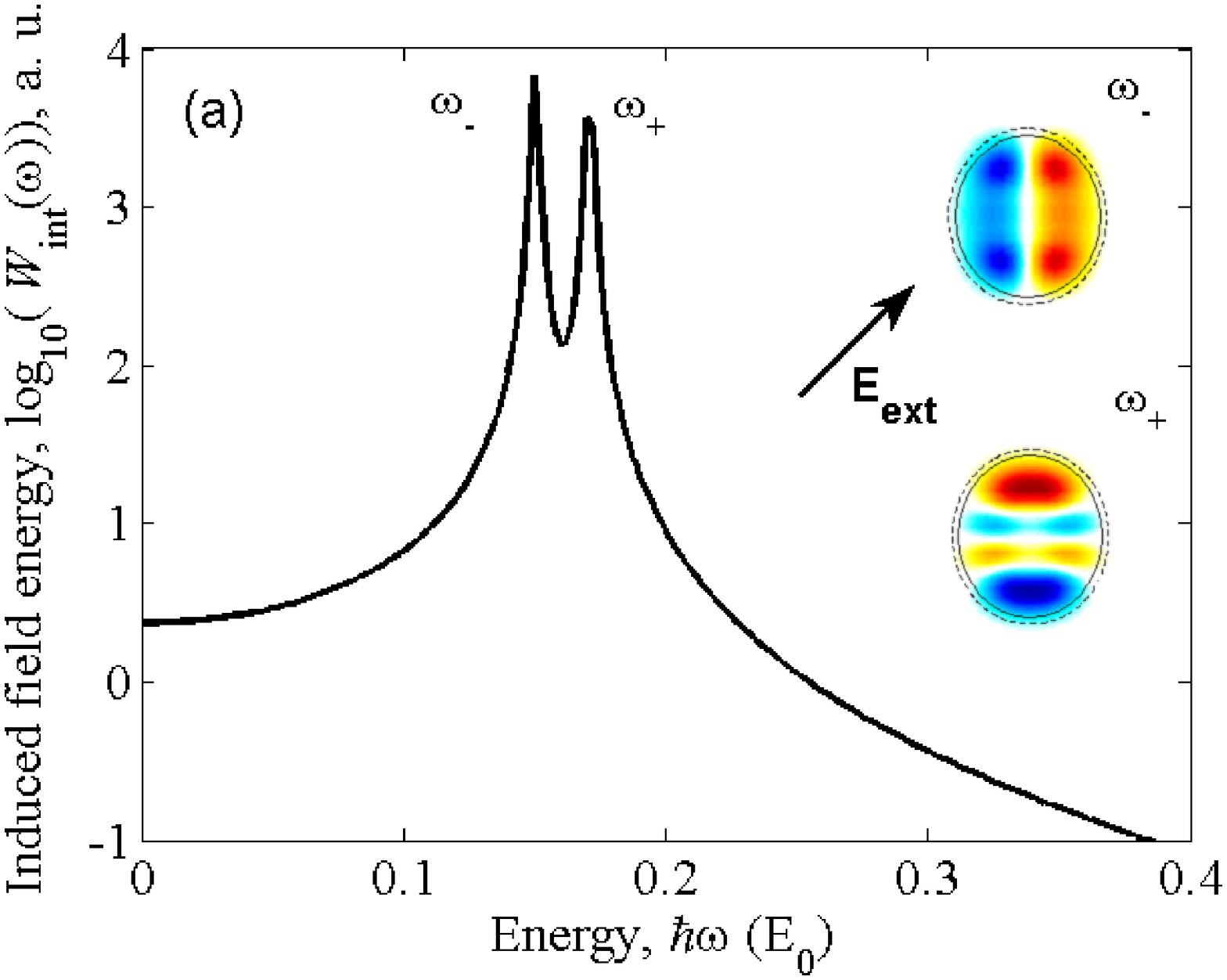}
\vspace{.3cm}
\includegraphics[width=5.cm,angle=270, bb=50 71 550 735]{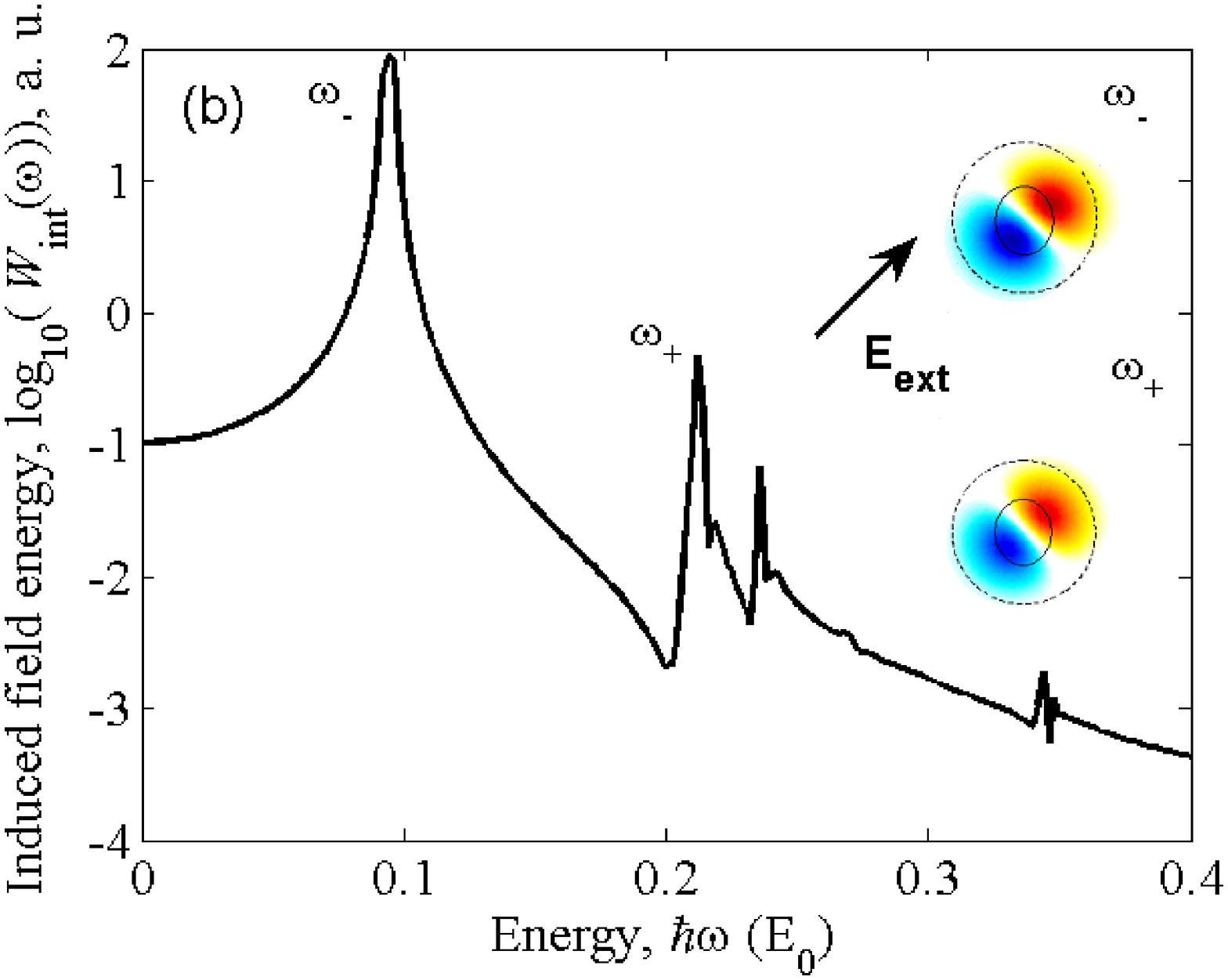}
\caption{\label{fig2} \footnotesize{ (a) Logarithm of the energy
of the induced electric field $(\log_{10}(W_{\text{ind}}))$ in an
elliptic rod of size $R=11$ $L$ with aspect ratio $a:b=1:1.3$ as a
function of photon energy $\hbar \omega$ of the external field in
units of $E_0$. Carrier density $\rho=(1/480)$ $L^{-3}$ and
$\gamma=10^{-3}$ $E_0$. The direction of the external field is
indicated by the arrow. Inset: induced charge density at the
resonant frequencies $\omega_+$ and $\omega_-$. The boundary of
the rod is shown using the solid line, and the dotted line shows
the set of classical turning points, corresponding to the positive
background potential.
 (b) Same as (a), but for size $R=3$ $L$.}}
\end{figure}
%\vspace{-1.5cm}
 The physics determining the value of $R_{\text {c}}$ may be illustrated
  considering an infinitely long cylindrical rod of radius $R$ and electron
density $\rho$. In the classical regime, the observed collective
oscillations are only weakly damped, indicating that they are well
separated from the quasi-continuum of single-particle excitations.
This leads to the condition $\omega_{\text {p}} R/v_{\text {F}}\gg
1$, stating that the plasmon phase velocity is greater
 than the Fermi velocity  $v_{\text {F}}$ of the electrons \cite{pines}.   For $r<R$ the electrons are trapped in a harmonic
potential  due to the uniform positive background \cite{kresin1}
and the characteristic collective frequency is $\omega_{\text
{p}}=\sqrt{4\pi e^2 \rho/(2 m_{\text {e}})}$. Estimating the Fermi
velocity using a bulk value $v_{\text F}=(3 \pi^2 \rho)^{1/3}\hbar
/m_{\text e}$ one obtains $R\gg R_{\text c}= \frac{
\pi^{1/6}3^{1/3}}{\sqrt{2}}\frac{\hbar}{e m_{\text e}^{1/2}
\rho^{1/6}} \approx 3.4$ $L$ for an electron density
$\rho=(1/480)$ $L^{-3}$,  which is in reasonable, but only
approximate, agreement with the calculated threshold $R_{\text
c}\approx 6.5$ $L$ shown in Fig. $1$.

For system sizes below $R_{\text {c}}$ the dominant excitations
are observed to shift towards lower energies as the positive
background potential becomes increasingly anharmonic. The
harmonicity criterion for excitations can be expressed as
$\sigma_0/R\le 1$, where $\sigma_0=\big(\frac{\hbar}{m_{\text e}
\omega_{\text p}}\big)^{1/2}$ is the classical turning point for
an electron in the ground state of the harmonic potential. This is
equivalent to the condition $R\ge R_{\text {c}}=(\frac{\hbar^2}{2
m_e \rho e^2})^{1/4}$, yielding $R \ge 3$ $L$.

The nature of the excitations changes as the characteristic size
$R$ crosses the quantum threshold $R_{\text c}$. In Fig.
\ref{fig2}, we show logarithm of the calculated  energy of the
induced field $(\log_{10}(W_{\text{ind}}))$ as a function of
external field frequency. As shown in Fig. 2(a), for relatively
large rod sizes,  the two distinct plasmon resonances labelled
$\omega_-$ and $\omega_+$ correspond to two orthogonal bipolar
charge distributions, as indicated in the inset. The spatial
orientations of these induced resonances are aligned with the
semi-major and semi-minor axes, and do not depend on the direction
of the incident field. In contrast, for system sizes less than
$R_{\text c}$ (Fig. \ref{fig2}(b)), the excitation spectrum
consists of several lower-intensity modes, dominated by a low
frequency resonance. However, the spatial characteristics of these
modes are different from the classical limit, i.e. they are {\it
aligned} with the incident field, indicating that Mie theory
breaks down in the quantum regime.

An advantage of the  self-consistent non-local response theory
described in this work is that it extends naturally to
inhomogeneous structures with nano-scale and atomically sharp
edges and corners. This is of great practical interest to
nano-photonic applications since such features are commonly
associated with substantial local field enhancements \cite{sers}.
However, the intuition driving such expectations is usually
derived from classical continuum field theory which may not be
applicable in this regime. When investigating  structures at the
microscopic level one needs to account quantitatively for changes
in the response spectrum. One would also like to describe the
breakdown of the continuum picture and quantum-mechanical
discretization effects which may ultimately lead to new and
interesting functionalities that are accessible by nano-scale
engineering.
%\hspace{3.cm}
\begin{figure}
\vspace{0.cm}
%\vspace{1.cm}
\includegraphics[width=9cm,angle=-90, bb= 45 40 580 555]{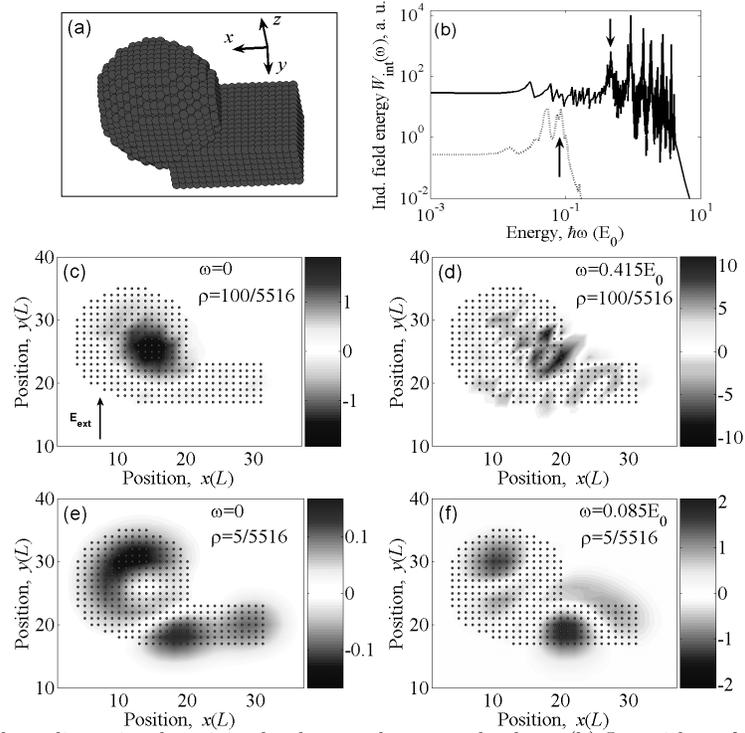}
%\vspace{1.5cm}
 \caption{\label{fig3} \footnotesize {(a) Illustration of three-dimensional  conjoined sphere  and
rectangular bar. (b) Logarithm of the induced field energy in the
structure for $N_{\text{el}}=100$ (solid line) and
$N_{\text{el}}=5$ (dotted line) as a function of the external
field frequency.  The material volume is  $V_{\text{mat}}=5516$
$L^{3}$. The arrows in the figure show the frequencies chosen for
results of calculations shown in (d) and (f). Induced charge
density for a nanoscale asymmetric structure for the indicated
charge densities and frequencies of the external field.
Temperature $T=0$ K and $\gamma=10^{-3}$ $E_0$. (c)
$N_{\text{el}}=100$, $\hbar \omega = 0$, the direction of the
external field is indicated by the arrow; (d) $N_{\text{el}}=100$,
$\hbar \omega = 0.415$ $E_0$. (e) $N_{\text{el}}=5$, $\hbar \omega
= 0$. (f) $N_{\text{el}}=5$, $\hbar \omega = 0.085$ $E_0$. }}
\end{figure}
\begin{figure}[t]
\vspace{0.cm}
\includegraphics[width=6.5cm,angle=270, bb= 33 33 542 712]{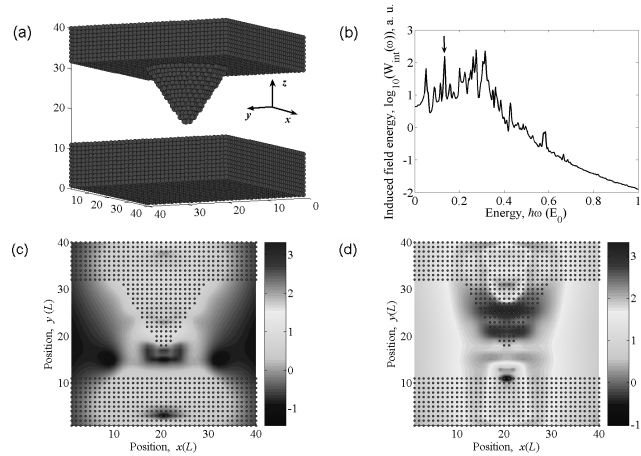}
%\vspace{-3. cm}
\vspace{0.cm} \caption{\label{fig4}\footnotesize { (a)
Three-dimensional image of a structure consisting of a sharp tip
and two parallel flat plates of material volume
$V_{\text{mat}}=33404$ $L^{3}$ containing $N_{\text{el}}=207$
electrons. The incident electric field is along the $(0,0,1)$
direction, temperature $T=0$ K and $\gamma=10^{-3}$ $E_0$.  (b)
Logarithm of the induced electric field energy in the structure as
a function of the field frequency,  the arrow shows the frequency
chosen for results of calculations shown in (d). (c) Logarithm of
the intensity of the induced electric field $(\log_{10}(|{\bf
E}_{\text{ ind}}({\bf{r}})|^2))$ in static limit, $\hbar
\omega=0$. (d) At resonance, $\hbar \omega=0.135$ $E_0$.}}
\end{figure}

To explore this we consider a representative asymmetric
nanostructured system of material volume $V_{\text{mat}}=5516$
$L^{3}$ consisting of a conjoined sphere and rectangular bar (Fig.
\ref{fig3}(a)) whose response strongly depends on the carrier
concentration (Fig. \ref{fig3}(b)). Plots of induced charge
distributions in response to an external electric field are shown
for different carrier concentrations  and frequencies of the
incident radiation.

For $N_{\text{el}}=100$ in the low-frequency limit (Fig.
\ref{fig3}(c)), this leads to a dipole-like anisotropic response
for which the induced field is not collinear to the external field
${\bf E}_{\text {ext}}$ (indicated by the arrow). Note,  that in
this case  the induced charge is localized within the boundaries
of the nanostructure, and follows
 the nanostructure's physical bounds given by the positive charge distribution,
 in agreement with expectations from classical field theory. At this relatively high carrier
concentration and at a high external field frequency  of
$\omega=0.415$ $E_0$, one observes a complex local enhancement of
the induced charge density (Fig. \ref{fig3}(d)).
 For the low carrier concentration ($N_{\text{el}}=5$), shown in Fig. \ref{fig3}(e) and
 (f) the induced charge density arises from  excitation of low
energy eigenstates  and the overall response is weaker. Note, the
induced charge is not localized within the boundaries of the
nanostructure. Varying the carrier concentration hence acts as a
``switch" that can activate different resonant regions in a
broken-symmetry atomic-scale structure.

At higher frequencies, the system response can be even more
complex. For example, in the case of low carrier concentrations
and resonance frequency $\omega=0.085$ $E_0$, the induced dipole
field is rotated with respect to the static limit (Fig.
\ref{fig3}(f)). Hence, by tuning the frequency of the external
field, discrete quantum states within anisotropic nanostructures
can either be accessed or avoided. This control exposes various
quantum functionalities that are beyond conventional Mie theory.

As an example, let us consider the induced electric field at the
interface between a sharp tip and a flat surface illuminated by
plane-wave radiation incident in the $z$ direction (Fig.
\ref{fig4}(a)). The total material volume of the system is
$V_{\text{mat}}=33404$ $L^{3}$.

 In the static limit, a moderate enhancement of field intensity is found
in close proximity to the tip.  Results of calculating logarithm
of the induced field intensity $(\log_{10}(|{\bf E}_{\text{
ind}}({\bf{r}})|^2))$ are shown in Fig. \ref{fig4}(c). While most
of the induced charge density accumulates at the surface, there is
considerable penetration into the bulk. In contrast, when the same
structure is illuminated at resonance, the induced plasmonic
response is much more intense than the low frequency limit. Here,
the induced field intensity  increases by $2$ orders of magnitude,
while the spatial dispersion of the ``hot spot'', as measured by
full width half maximum, remains approximately the same (see Fig.
\ref{fig4}(d)). This demonstrates that custom-designed
atomic-scale tip structures may be used to control
 the near field and fully quantum-mechanical modeling can quantitatively account for local enhancements, as well as
screening and focusing effects.

%Version 2\\
In summary, a non-local response theory describing the interaction
of electromagnetic radiation with inhomogeneous nano-scale
structures has been developed and implemented. This approach may
be used to describe metals and semiconductors in the metallic
regime. For simple geometries semi-empirical continuum-field Mie
theory is found to break down beyond a critical coarseness. At
this level of coarse graining, the response is no longer
exclusively determined by particle shape. In more complex
inhomogeneous geometries excitations of quantum and classical
character can coexist. For objects with nano-scale sharp features,
our real-space response theory uncovers new functionalities, such
as local resonances that are activated by tuning the carrier
concentration or the frequency of the incident field.

 We acknowledge discussions with P. B.  Littlewood and V. Kresin. This work is supported by DARPA.
Computational facilities have been generously provided by the HPC Center at USC.

%\end{multicols}
\end{document}